\documentclass[aps,prl,twocolumn,superscriptaddress,letterpaper]{revtex4-2}

\usepackage{graphicx}
\usepackage{bm}
\usepackage{amssymb}
\usepackage{color}
\usepackage{amsmath}
\usepackage{ulem}
\usepackage[colorlinks=true,linkcolor=blue,anchorcolor=black,citecolor=blue,filecolor=black,menucolor=black,runcolor=black,urlcolor=blue]{hyperref}

\begin{document}

\title{Anomalous Floquet non-Hermitian skin effect in a ring resonator lattice}

\author{He Gao}
\affiliation{Department of Mechanical Engineering, The Hong Kong Polytechnic University, Kowloon, Hong Kong SAR, China}

\author{Haoran Xue}
\email{haoran001@e.ntu.edu.sg}
\affiliation{Division of Physics and Applied Physics, School of Physical and Mathematical Sciences, Nanyang Technological University,
Singapore 637371, Singapore}

\author{Zhongming Gu}
\affiliation{Institute of Acoustics, School of Physics Science and Engineering, Tongji University, Shanghai 200092, China}

\author{Linhu Li}
\affiliation{Guangdong Provincial Key Laboratory of Quantum Metrology and Sensing \& School of Physics and Astronomy, Sun Yat-Sen University (Zhuhai Campus), Zhuhai 519082, China}

\author{Weiwei Zhu}
\affiliation{Department of Physics, National University of Singapore, Singapore 117542, Singapore}

\author{Zhongqing Su}
\affiliation{Department of Mechanical Engineering, The Hong Kong Polytechnic University, Kowloon, Hong Kong SAR, China}

\author{Jie Zhu}
\email{jiezhu@tongji.edu.cn}
\affiliation{Institute of Acoustics, School of Physics Science and Engineering, Tongji University, Shanghai 200092, China}

\author{Baile Zhang}
\email{blzhang@ntu.edu.sg}
\affiliation{Division of Physics and Applied Physics, School of Physical and Mathematical Sciences, Nanyang Technological University,
Singapore 637371, Singapore}
\affiliation{Centre for Disruptive Photonic Technologies, Nanyang Technological University, Singapore, 637371, Singapore}

\author{Y.D. Chong}
\email{yidong@ntu.edu.sg}
\affiliation{Division of Physics and Applied Physics, School of Physical and Mathematical Sciences, Nanyang Technological University,
Singapore 637371, Singapore}
\affiliation{Centre for Disruptive Photonic Technologies, Nanyang Technological University, Singapore, 637371, Singapore}

\begin{abstract}
We present a one-dimensional coupled ring resonator lattice exhibiting a variant of the non-Hermitian skin effect (NHSE) that we call the anomalous Floquet NHSE.  Unlike existing approaches to achieving the NHSE by engineering gain and loss on different ring segments, our design uses fixed on-site gain or loss in each ring.  The anomalous Floquet NHSE is marked by the existence of skin modes at every value of the Floquet quasienergy, allowing for broadband asymmetric transmission.  Varying the gain/loss induces a non-Hermitian topological phase transition, reversing the localization direction of the skin modes. An experimental implementation in an acoustic lattice yields good agreement with theoretical predictions, with a very broad relative bandwidth of around 40\%.  
\end{abstract}

\maketitle

\textit{Introduction.}---Non-Hermitian systems can exhibit a range of striking phenomena with no counterparts in the Hermitian regime \cite{feng2017, el2018, miri2019, konotop2016, bergholtz2021}.  The one that has perhaps attracted the most interest over the past two decades is parity-time (PT) symmetry, which allows a non-Hermitian system to host real eigenvalues and non-Hermitian phase transitions \cite{bender1998}; this has been realized and extensively investigated in photonics, and other platforms such as acoustics, using classical gain and/or loss to implement non-Hermiticity \cite{guo2009, ruter2010, hodaei2014, feng2014}.  Recently, much attention has been drawn to another non-Hermitian phenomenon called the non-Hermitian skin effect (NHSE) \cite{alvarez2018, xiong2018, gong2018, yao2018, kunst2018}, whereby a non-Hermitian lattice hosts an extensive number of boundary-localized eigenmodes called skin modes. The NHSE is theoretically intriguing as it signifies a breakdown of Bloch's theorem for non-Hermitian systems \cite{lee2016, alvarez2018, xiong2018, gong2018, yao2018, yao2018b, kunst2018, lee2019} and is associated with novel non-Hermitian bulk-boundary correspondences based on the windings of complex energy spectra \cite{gong2018, okuma2020, zhang2020, borgnia2020non}; moreover, it may have application possibilities in sensing \cite{budich2020, mcdonald2020} and lasing \cite{Longhi2018, bofeng2022}.  Over the past two years, experimental realizations of the NHSE have rapidly emerged in electric circuits \cite{helbig2020, liu2021, zou2021observation}, quantum walks \cite{xiao2020, xiao2021observation}, phononic metamaterials \cite{brandenbourger2019, ghatak2020, chen2021realization, zhang2021observation, zhang2021acoustic}, optical fiber-based synthetic lattices \cite{weidemann2020, wang2021}, and active particle systems \cite{palacios2021guided}.  A notable obstacle to such experimental studies is the fact that the simplest theories of the NHSE (though not all of them \cite{yi2020, zhang2022universal}) involve tight-binding models with nonreciprocal inter-site couplings, which tend to be difficult to implement.

In this work, we study an interesting form of the NHSE that arises in lattices of coupled ring resonators.  We theoretically analyze and experimentally implement one-dimensional lattices of coupled ring resonators with on-site gain and/or loss, in a configuration that does not require placing gain/loss on different segments of some rings (which had been the approach adopted in earlier works to induce nonreciprocal inter-site couplings \cite{longhi2015, zhu2020, zhang2021observation, lin2021, liu2021supermetal}).  Coupled-ring lattices have two key features relevant to the NHSE: (i) if waves experience negligible back-reflection while propagating in the lattice  (a standard assumption), certain lattice configurations allow the eigenmodes to split into two circulation sectors that are each effectively nonreciprocal \cite{hafezi2013, liang2013, Pasek2014, leykam2018}; (ii) in the strong-coupling regime, eigenmodes are solutions to a Floquet, rather than Hamiltonian, eigenproblem, and can deviate qualitatively from tight-binding models \cite{rudner2013}.

We show that our coupled-ring lattice exhibits an ``anomalous Floquet NHSE'' whereby the real part of the Floquet quasienergy winds through its entire $2\pi$ range---something that cannot occur in Hamiltonian models \cite{lee2016, alvarez2018, xiong2018, gong2018, yao2018, yao2018b, kunst2018, lee2019}.  The skin modes manifest in the \textit{entire} range of Floquet quasienergies, rather than a small interval corresponding to the bulk energy band. This behavior can potentially be used to achieve asymmetric transmission over a much broadband bandwidth than in non-Floquet systems.  As a proof of principle, we construct an acoustic lattice exhibiting the anomalous Floquet NHSE, and demonstrate experimental results that are consistent with the theoretical model, and in quantitative agreement with full-wave simulations, over a $\sim 40\%$ relative frequency bandwidth.  This design may also be useful for achieving broadband NHSE, and related phenomena, in platforms such as microwave photonics \cite{hu2015, gao2016} and nanophotonics \cite{hafezi2013, zhao2019, mittal2019, afzal2020}.

\begin{figure*}
  \centering
  \includegraphics[width=0.9\textwidth]{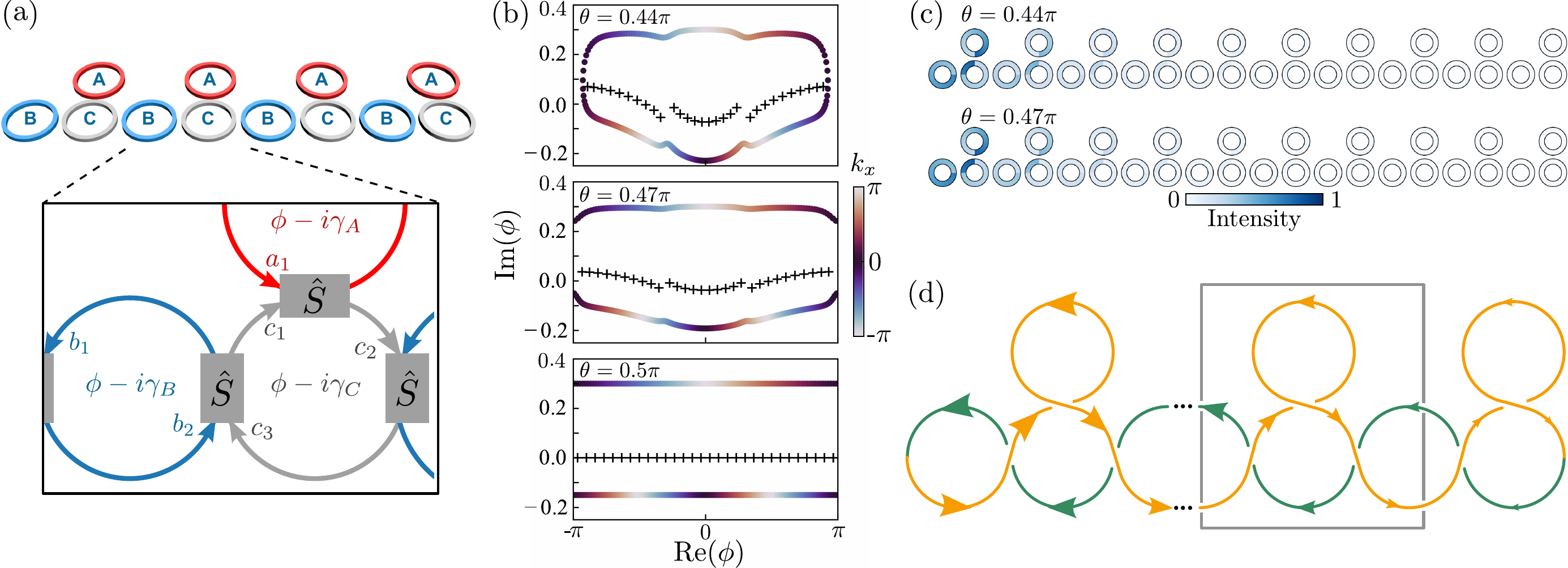}
  \caption{(a) Schematic of a finite lattice of coupled ring resonators (top) and a close-up view (bottom).  One round trip in each ring multiplies the wave amplitude by $\exp[i(\phi - i \gamma_{A,B,C})]$.  The phase shift $\phi$ is interpreted as a Floquet quasienergy.  The wave coefficients at six points in the unit cell are denoted by $(a_1,b_1,b_2,c_1,c_2,c_3)$.  (b) Complex quasienergy spectrum under PBC (colored dots) and OBC with a lattice of 10 unit cells (black crosses) for $\theta \in \{0.44\pi, 0.47\pi, 0.5\pi\}$.  (c) Intensity profiles, summed over all eigenmodes, for the lattices with OBC and $\theta\in\{0.44\pi, 0.47\pi\}$. Both of these cases host skin modes, despite the qualitative differences in the PBC spectra. (d) Wave propagation patterns under perfect coupling, $\theta = 0.5\pi$.  The grey rectangle marks one unit cell.  If an OBC mode's right-moving path (orange) is damped, then its left-moving path (green) is amplified and the mode profile is exponentially localized to the left, as indicated by the sizes of the arrows.   In (b)--(c), the gain/loss parameters are $\gamma_A=-0.6$ and $\gamma_B=\gamma_C=0.3$.}
  \label{fig1}
\end{figure*}

\textit{Theory.}---Consider the one-dimensional lattice shown in Fig.~\ref{fig1}(a), with each unit cell consisting of three distinct coupled ring resonators.  Waves propagating in each ring acquire phase shifts as well as experiencing amplitude growth/decay due to gain/loss.  For the rings drawn in red, blue, and grey in Fig.~\ref{fig1}(a), and labelled $\{A, B, C\}$, one round-trip multiplies the complex wave amplitude by $\exp[i(\phi - i \gamma_{A,B,C})]$.  If $\phi$ is real, it corresponds to the round-trip phase shift in each ring, while positive (negative) values of $\gamma_{A,B,C}$ correspond to gain (loss) in the respective rings.  We consider only one of the two choices of circulation directions in the lattice---say, counterclockwise in the $A$ and $B$ rings, and clockwise in the $C$ rings---with no ``spin-flipping'' \cite{hafezi2013}.  This restriction to one circulation sector effectively breaks reciprocity \cite{hafezi2013, liang2013, Pasek2014, leykam2018}.  For simplicity, we let all rings share the same $\phi$ parameter, which is interpreted as a Floquet quasienergy \cite{liang2013, Pasek2014}.  However, the behaviors discussed below generalize to rings of different $\phi$; as an example, in the Supplemental Materials we describe the case where the $A$ and $B$ rings are anti-resonant with the $C$ rings \cite{SM}.  Note also that each ring is assumed to have uniform gain or loss (or neither), unlike previously-studied non-Hermitian coupled-ring models that have gain or loss on different parts of each ring \cite{longhi2015, zhu2020, zhang2021observation, lin2021}.

Wave propagation within the lattice can be modelled with the transfer matrix method \cite{hafezi2013, liang2013, Pasek2014, leykam2018}.  As shown in Fig.~\ref{fig1}(a), we let $\{a_1,b_1,b_2,c_1,c_2,c_3\}$ denote the complex wave amplitudes at six points in the unit cell, just before the coupling regions between adjacent rings.  The coupling is described by a scattering matrix $\hat{S} = \cos(\theta) - i \hat{\sigma}_1 \sin\theta$, where $\hat{\sigma}_1$ is a Pauli matrix and the angle $\theta$ describes the coupling strength, with $\theta = \pi/2$ corresponding to perfect coupling.

For a given set of source-free boundary conditions, such as periodic boundary conditions (PBC) or open boundary conditions (OBC), the scattering and coupling equations can be cast as an Floquet eigenproblem of the form $\hat{U} |\Psi\rangle = e^{-i\phi} |\Psi\rangle$, where $\hat{U}$ is an evolution operator and $|\Psi\rangle$ is a vector of complex wave amplitudes \cite{hafezi2013, liang2013, Pasek2014, SM}.  Physically, $\mathrm{Re}[\phi]$ describes the round trip phase shifts in each ring, and $\mathrm{Im}[\phi]$ is an additional gain or loss applied to each ring (on top of $\gamma_{A,B,C}$) to produce a self-consistent wave pattern in the lattice.  In a real system, $\phi$ is typically proportional to the operating frequency, though the other parameters $\gamma_{A,B,C}$ and $\theta$ may also vary with frequency.

Figure~\ref{fig1}(b) shows the complex quasienergy spectra for $\gamma_A=-0.6$, $\gamma_B=\gamma_C=0.3$, and different values of $\theta$.  In each plot, the colored markers correspond to PBC, and the black crosses correspond to OBC for a lattice of 10 unit cells with the same termination conditions as in Fig.~\ref{fig1}(a).  For $\theta = 0.44\pi$ [upper panel of Fig.~\ref{fig1}(b)], the behavior is similar to previous non-Hermitian Hamiltonian models exhibiting the NHSE \cite{lee2016, alvarez2018, xiong2018, gong2018, yao2018, yao2018b, kunst2018, lee2019}: the quasienergies for PBC form a loop with nonzero point gap winding, while the quasienergies for OBC form an arc enclosed by and connecting to the loop. The OBC eigenmodes exhibit the NHSE, as shown in the upper panel of Fig.~\ref{fig1}(c).

With a larger coupling angle, $\theta = 0.47\pi$, the complex quasienergy spectrum is qualitatively different.  As shown in the middle panel of Fig.~\ref{fig1}(b), under both PBC and OBC, the quasienergies wrap across the entire range of $\mathrm{Re}[\phi] \in [-\pi, \pi]$, and the arc of OBC quasienergies never meets the PBC quasienergies.  Nonetheless, in this regime there exists a macroscopic number of skin modes, as shown in the lower panel of Fig.~\ref{fig1}(c).

This type of complex spectrum is made possible by the fact that the present model is a Floquet system---i.e., it is governed by a (non-unitary) evolution operator $\hat{U}$ rather than a (non-Hermitian) Hamiltonian, so that $\mathrm{Re}[\phi]$ is an angle variable \cite{rudner2013,Pasek2014}.  In other Floquet models, similar quasienergy wrappings have been shown to give rise to so-called anomalous Floquet insulators \cite{rudner2013, hu2015, gao2016, Fleury2021} and other anomalous Floquet phases \cite{huang2020, zhu2022time}.  Similarly, we call the present behavior the ``anomalous Floquet NHSE''.

Because the skin modes have quasienergies spanning the entire range of $\mathrm{Re}[\phi] \in [-\pi, \pi]$, the anomalous Floquet NHSE is a broadband phenomenon.  Instead of using the standard definition of point gap winding \cite{gong2018, okuma2020, zhang2020, borgnia2020non}, it can be characterized by \cite{SM}
\begin{equation}
  \nu(\phi_r) = \int_{\text{BZ}}\frac{dk_x}{2\pi}\frac{d}{dk_x}
  \arg\det[\hat{U}-e^{-i\phi_r}],
  \label{winding-number}
\end{equation}
which describes the winding of $e^{-i\phi}$ (not $\phi$) around a reference point $e^{-i\phi_r}$.  All three cases in Fig.~\ref{fig1}(b) have $\nu(0)=1$, consistent with the occurrence of skin modes.

The physical origin of the anomalous Floquet NHSE can be understood using the extremal case of $\theta = 0.5\pi$, or perfect inter-ring coupling.  As shown in the bottom plot of Fig.~\ref{fig1}(b), under PBC there are two quasienergy bands with constant $\mathrm{Im}[\phi]$, whereas all the OBC modes have the same $\mathrm{Im}[\phi]$.  Perfect inter-ring coupling causes wave trajectories to form two distinct paths in the bulk, as shown in Fig.~\ref{fig1}(d).  Under PBC, the paths are completely decoupled; for the right-moving path, the wave amplitudes are multiplied by $\exp[i(2\phi - i\Gamma_R)]$ per unit cell, where $\Gamma_R = \gamma_A + (\gamma_B + \gamma_C)/2$, and applying the Bloch condition yields the dispersion relation $\phi = k/2 + i\Gamma_R/2$, where $k$ is the Bloch wavenumber with the lattice period normalized to unity.  The left-moving wave amplitudes are multiplied by $\exp[i(\phi - i\Gamma_L)]$ per unit cell, where $\Gamma_L = (\gamma_B + \gamma_C)/2$, so their dispersion relation is $\phi = -k + i\Gamma_L$.  Under OBC, the two paths meet at the ends of the lattice, as seen in Fig.~\ref{fig1}(d).  A self-consistent wave pattern must have no net gain/loss after each round trip, so $\mathrm{Im}[\phi] = (\Gamma_L + \Gamma_R) / 3 = (\gamma_A + \gamma_B + \gamma_C) / 3$, consistent with the complex quasienergy spectrum shown in Fig.~\ref{fig1}(b).  If the left-moving/right-moving path in an OBC mode is amplified/damped, the mode profile grows exponentially to the left, as indicated by the arrow sizes in Fig.~\ref{fig1}(d). Such a formation picture of the skin modes under OBC applies to all $\phi$ in the perfect coupling case, thus leading to the anomalous Floquet NHSE.

\begin{figure}
  \centering
  \includegraphics[width=\columnwidth]{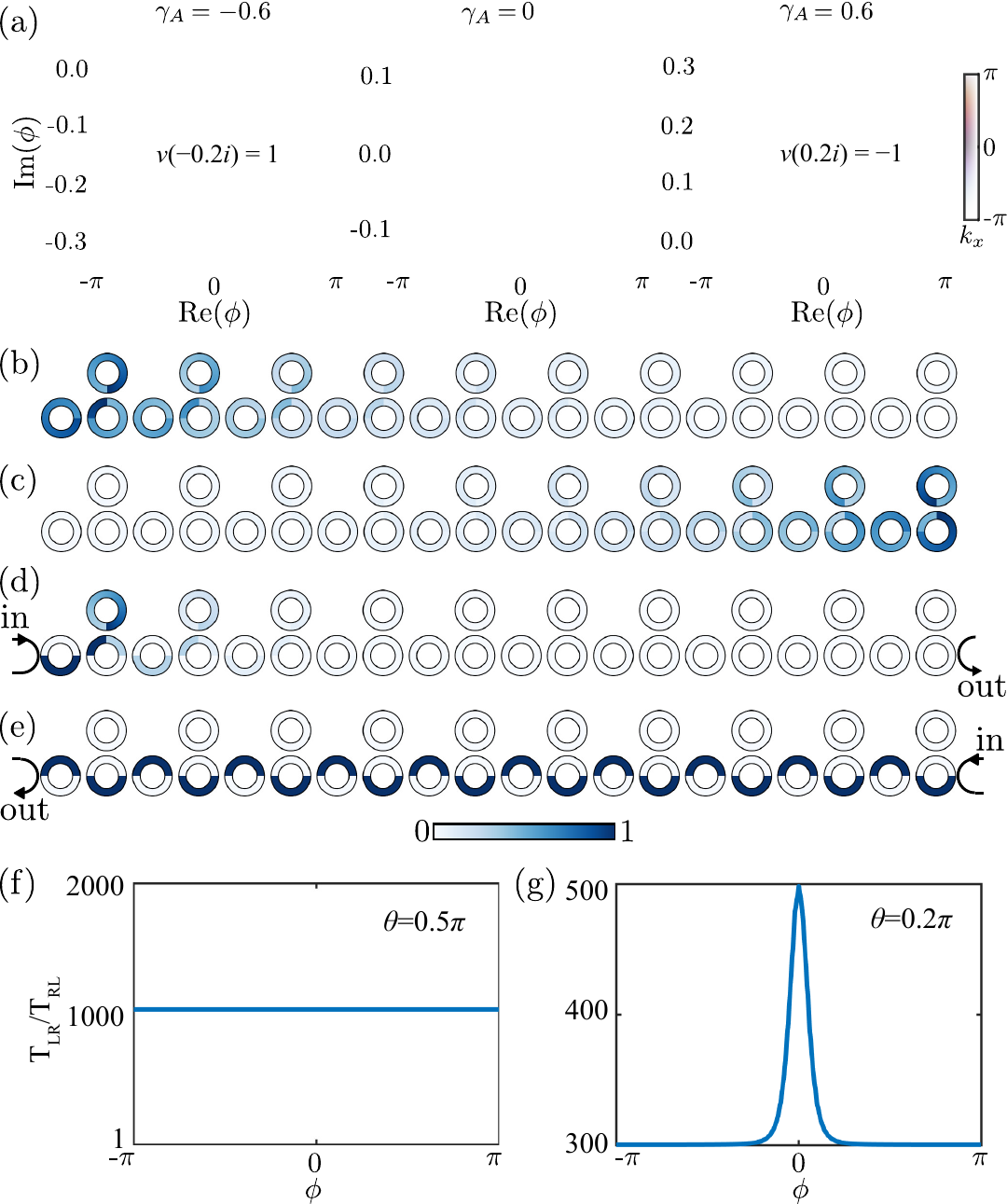}
  \caption{(a) Spectra for a lattice under PBC with gain and loss parameters: $\gamma_B=\gamma_C=0$, $\gamma_A=-0.6, 0, 0.6$ (left to right). (b)--(c) Profile of a typical skin mode in a lattice under OBC, with $\gamma_B=\gamma_C=0$, and (b) $\gamma_A=-0.6$, (c) $\gamma_A=0.6$. (d)--(e) Field distributions when exciting the lattice from the left and the right, respectively. (f)--(g) Plots of transmission ratio $\mathrm{T_{LR}/T_{RL}}$ against $\phi$ for $\theta=0.5\pi$ and $\theta=0.2\pi$, respectively. Here $\mathrm{T_{LR}}$ ($\mathrm{T_{RL}}$) denotes transmission from right to left (left to right). In (d)--(e), the gain/loss parameters are the same as in (a) and the input/output couplings are $\theta_{\mathrm{in/out}}=0.5\pi$.}
  \label{fig2}
\end{figure}

We have chosen a coupling strength and gain/loss configuration that is feasible for experiments (e.g, rings in the same row have the same gain/loss).  For other parameter choices, we observe similar behaviors, including other instances of the anomalous Floquet NHSE.  The PBC spectrum can even exhibit multiple loops with different winding directions, resulting in skin modes localized at different boundaries, consistent with the above bulk-boundary correspondence principle \cite{SM}.

\begin{figure*}
  \centering
  \includegraphics[width=0.8\textwidth]{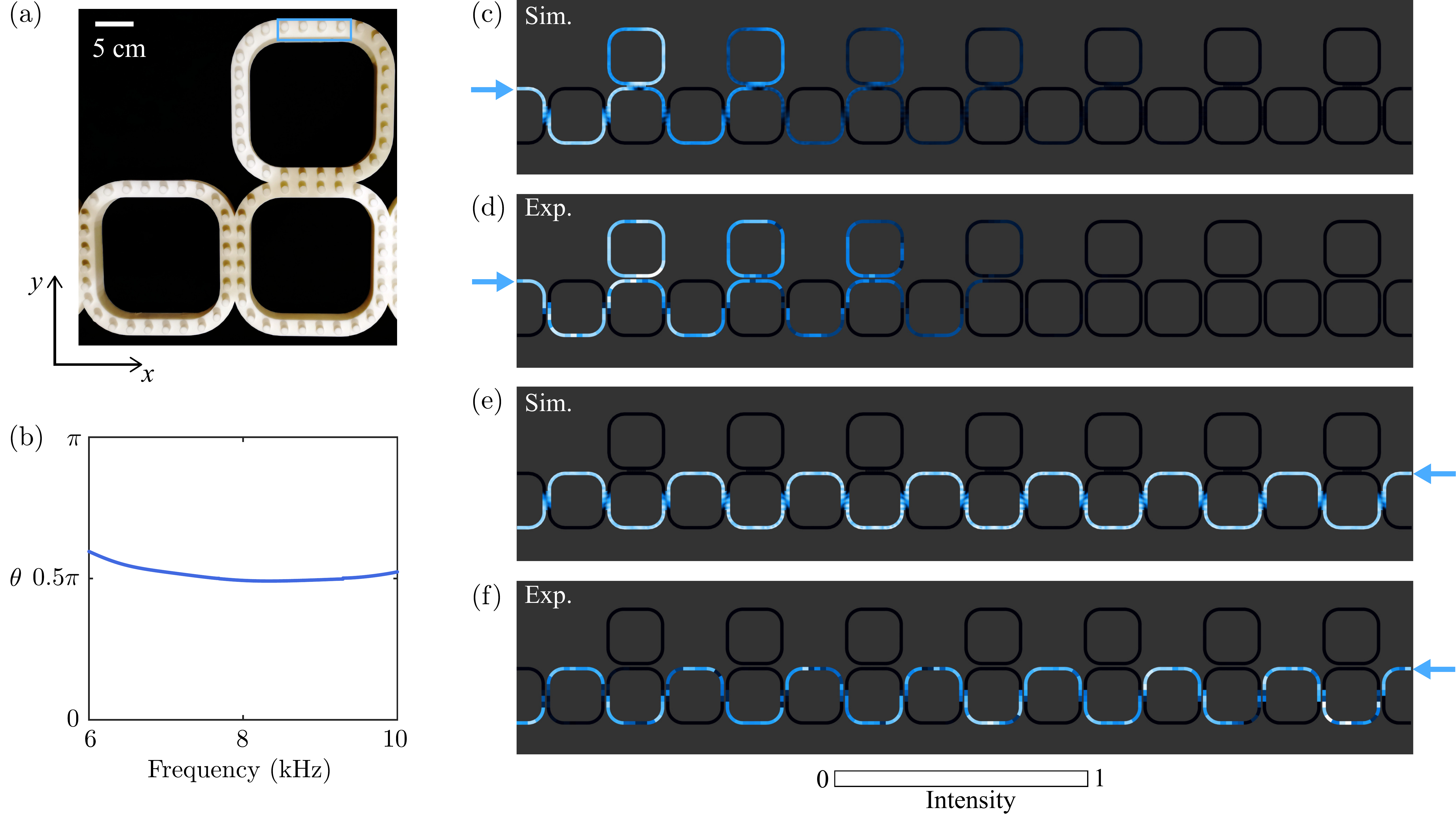}
  \caption{(a) Photograph of one unit cell in the experimental sample. (b) Plot of effective coupling angle $\theta$ against frequency, extracted from full-wave numerical simulations as described in the Supplemental Materials \cite{SM}. (c)--(f) Full-wave acoustic simulations and experimental measurements of the acoustic intensity under: (c),(d) left incidence and (e),(f) right incidence. The excitation frequency is 8000 Hz. In (d) and (f), we multiply the measurement results by a factor of $\exp(\alpha l)$, where $\alpha$ is a decay factor and $l$ is the path length, to compensate for background losses \cite{SM}.  In the full-wave simulations, intrinsic losses are absent and the additional loss in the $A$ ring is implemented as a nonzero imaginary part in the sound speed. The experiment and simulations can be described by the Floquet model with $\gamma_A=-0.35$, $\gamma_B=0$ and $\gamma_C=0$.}
  \label{fig3}
\end{figure*}

Moreover, gain and loss can induce a topological transition whereby the Floquet point gap windings reverse direction, and the skin modes correspondingly switch boundaries. To demonstrate this, consider a lattice with the following parameters: $\gamma_B=\gamma_C=0$ and $\theta=0.5\pi$. In addition, we introduce a variable $\gamma_A$ that can be tuned to achieve the topological transition. Figure~\ref{fig2}(a) shows the spectra under PBC as $\gamma_A$ is increased from $-0.6$ to $0.6$.  For $\gamma_A=-0.6$, the PBC spectrum consists of two branches, as shown in the left panel of Fig.~\ref{fig2}(a).  We find $\nu = 1$ (for $\phi_r = -0.2i$), and the skin modes are localized to the left boundary as shown in Fig.~\ref{fig2}(b).  Upon increasing $\gamma_A$, the two branches of PBC quasienergies approach each other, meeting at $\gamma_A=0$ as shown in the middle panel of Fig. \ref{fig2}(a). This is the phase transition point.  As $\gamma_A$ is further increased, the PBC spectrum again splits into two branches with $\nu=-1$ (for $\phi_r = 0.2i$), as shown in the right panel of Fig.~\ref{fig2}(a). The OBC modes are now skin modes localized on the right boundary, as shown in Fig.~\ref{fig2}(c).  Thus, our Floquet model enables the active control of the NHSE, which may be useful for reconfigurable devices based on optical platforms where gain and/or loss can be tuned by external pumping \cite{zhao2019}.

Thus far, we have investigated how the model behaves under source-free conditions.  In Figs.~\ref{fig2}(d)--(e), we study the transmission properties when inputs and outputs are attached to the ends of the lattice.  The lattice parameters are the same as in the left panel of Fig.~\ref{fig2}(a); the couplings between the couplers and the lattice are set to $0.5\pi$ (i.e., perfect coupling). As shown in Fig.~\ref{fig2}(d)--(e), the wave is continuously attenuated (unattenuated) when transmitted left-to-right (right-to-left).  This is consistent with the localization of the skin modes, and with the previously discussed differences in relative gain between left- and right-moving paths. Furthermore, such an asymmetric transmission behavior is broadband in the anomalous Floquet NHSE case [Fig.~\ref{fig2}(f)]. When the coupling is decreased and the skin modes only exist for a small range of $\phi$, the broadband property no longer holds [Fig.~\ref{fig2}(g)]. 

\textit{Experiment.}---We performed a proof-of-principle experimental demonstration using an acoustic lattice.  As shown in Fig.~\ref{fig3}(a), the lattice unit cell consists of three coupled rings with the same dimensions. Each ring is filled with air and surrounded by rigid walls, with adjacent rings connected by small channels (see Supplemental Material for detailed structural parameters \cite{SM}). The couplings are optimized so that the effective $\theta$, derived from full-wave simulations \cite{SM}, is approximately $0.5\pi$ over a wide range of frequencies, as shown in Fig.~\ref{fig3}(b).

Unlike the previously-discussed theoretical models, the experimental sample is purely lossy.  There are two sources of loss: the first is intrinsic loss due to material absorption, which is the same for all rings.  The second type of loss comes from the intentional introduction of absorbing materials to a segment of each $A$ ring, indicated by the blue box in Fig.~\ref{fig3}(a).

The lattice is excited by a monochromatic signal with a frequency of 8000 Hz incident from the left or right end, as indicated by the blue arrows in Figs.~\ref{fig3}(c)--(f).  We probe the acoustic intensities at various positions by inserting microphones into the rings, with a step of 28~mm (see Supplemental Material for more details on the measurement \cite{SM}).  For ease of visualization, the measured intensities in Figs.~\ref{fig3}(d) and (f) are compensated by multiplying the measured intensity by a factor $e^{\alpha l}$, where $\alpha$ is a decay factor induced by the background loss and $l$ is the propagation length \cite{SM}. The uncompensated intensity fields are shown in the Supplemental Material \cite{SM}.

When the signal is injected from the left, the acoustic intensity attentuates rapidly, as demonstrated in Figs.~\ref{fig3}(c)--(d).  By contrast, if the signal is injected from the right, the attenuation is negligible, as seen in Figs.~\ref{fig3}(e)--(f).  This is consistent with the behavior discussed in Figs.~\ref{fig2}(d)--(e). The asymmetric transmission is experimentally observed in a frequency window ranging from 6500 Hz to 10000 Hz (i.e., a relative bandwidth of $\sim 40\%$), consistent with the theoretical expectations that the anomalous Floquet NHSE should be broadband (experimental results at other frequencies are given in the Supplemental Material \cite{SM}).

\textit{Discussion.}---We have shown theoretically and experimentally that the NHSE can occur in a lattice of coupled ring resonators.  By analyzing lattice modes as Floquet eigenstates \cite{hafezi2013, liang2013, Pasek2014}, we found a new variant of the NHSE, the anomalous Floquet NHSE, in which skin modes exist at every quasienergy.  This phenomenon arises from the special features of Floquet bandstructures, which had previously been exploited in anomalous Floquet insulators to produce broadband and extraordinarily robust topological states \cite{Fleury2021}.  Similarly, the anomalous Floquet NHSE allows for broadband asymmetric transmission enabled by skin modes.

We note that there have been previous proposals to realize the non-anomalous NHSE in coupled-ring lattices, by designating certain rings as ``coupling rings'' with varying gain/loss on different arms \cite{longhi2015, zhu2020, zhang2021observation, lin2021}.  By comparison, our design assigns to each ring a certain level of gain/loss, and does not require gain/loss engineering on ring segments.  It may thus be easier to implement, especially on other platforms such as photonics.  There, our model may be helpful for realizing NHSE-aided lasing \cite{Longhi2018, bofeng2022}, and for studying the interplay of the NHSE with other non-Hermitian effects \cite{konotop2016} or nonlinearities \cite{smirnova2020, wimmer2015, lang2021non}.  The coupled-ring model can also be mapped to helical waveguide arrays \cite{leykam2016} or temporally modulated Floquet systems \cite{rudner2013}.  Finally, it would be interesting to generalize the lattice to two or higher dimensions \cite{SM}, in order to investigate higher-dimensional versions of the anomalous Floquet NHSE.

\acknowledgments{H.G. and H.X. contributed equally to this work. H.G., Z.G., Z.S. and J.Z. acknowledge support from the Research Grants Council of Hong Kong SAR (Grant Nos. 15205219, C6013-18G and AoE/P-502/20) and the National Natural Science Foundation of China (Grant No.11774297).  H.X., B.Z. and Y.C. acknowledge support from the Singapore MOE Academic Research Fund Tier 3 Grant MOE2016-T3-1-006, Tier 2 Grant MOE2019-T2-2-085, Tier 1 Grant RG148/20, and Singapore National Research Foundation (NRF) Competitive Research Program (CRP) (NRF-CRP23-2019-0007).}

\end{document}